\documentclass[preprint,showpacs,preprintnumbers,amsmath,amssymb]{revtex4}

\usepackage{graphicx}
\usepackage{dcolumn}
\usepackage{bm}

\begin{document}

\preprint{Submission to Phys. Rev. B}

\title{
Magnetic structure of Tb$_3$RuO$_7$ possessing 
alternating-bond spin-$\frac{3}{2}$ Ru chains 
in a nonuniform internal magnetic field: 
appearance of a partially disordered state 
}

\author{Masashi Hase$^1$}
 \email{HASE.Masashi@nims.go.jp}
\author{Andreas D\"onni$^2$}
\author{Vladimir Yu. Pomjakushin$^3$}

\affiliation{%
${}^{1}$Research Center for Advanced Measurement and Characterization, 
National Institute for Materials Science (NIMS), 
1-2-1 Sengen, Tsukuba, Ibaraki 305-0047, Japan \\
${}^{2}$International Center for Materials Nanoarchitectonics (WPI-MANA), 
National Institute for Materials Science (NIMS),
1-1 Namiki, Tsukuba, Ibaraki 305-0044, Japan \\
${}^{3}$Laboratory for Neutron Scattering and Imaging, Paul Scherrer Institut (PSI), 
CH-5232 Villigen PSI, Switzerland 
}%

\date{\today}

\begin{abstract}

We report on the magnetic structures of 
Tb$_3$RuO$_7$ and Nd$_3$RuO$_7$
determined from powder neutron-diffraction experiments. 
In Tb$_3$RuO$_7$, 
alternating-bond spin-$\frac{3}{2}$ Ru1-Ru2 chains are formed. 
The Ru1 and Ru2 moments are 
ordered and disordered (paramagnetic), respectively, below 15 K. 
This result indicates the 
appearance of a partially disordered (PD) state, 
although the two Ru sites are very similar to each other. 
In Nd$_3$RuO$_7$, 
the Ru1 and Ru2 ions form independent uniform chains. 
Both the Ru1 and Ru2 moments are ordered below 17.5 K. 
Probably, the main source of the PD state in Tb$_3$RuO$_7$ is 
the magnetic frustration in the exchange interactions. 
The internal magnetic field generated by the Tb ordered moments at the Ru1 sites 
is different from that at the Ru2 sites.
We speculate that 
the different (nonuniform) internal magnetic field enhances 
the difference in the properties between 
the Ru1 and Ru2 moments. 
We also report on the magnetic entropy changes of 
Tb$_3$RuO$_7$ and Gd$_3$RuO$_7$.  

\end{abstract}

\pacs{75.25.-j, 75.10.Pq, 75.47.Lx, 75.30.Sg}


\maketitle

\section{INTRODUCTION}


Spin chains 
to which a nonuniform magnetic field is applied are expected to exhibit exotic properties. 
For example, 
a staggered magnetic field is induced by 
a uniform external magnetic field at the spin chain with 
an alternating $g$ tensor and 
the antisymmetric interaction of the Dzyaloshinsky--Moriya (DM) type 
with an alternating $D$ vector. 
The staggered magnetic field is known to appear in real materials such as 
Cu(C$_6$H$_5$COO)$_2 \cdot$3H$_2$O (copper benzoate) \cite{Dender97,Nagata76,Asano00,Nojiri06}, 
CuCl$_2 \cdot$2[(CD$_3$)$_2$] \cite{Kenzelmann04,Kenzelmann05}, 
Yb$_4$As$_3$ \cite{Oshikawa99,Kohgi01},  
PM$\cdot$Cu(NO$_3$)$_2 \cdot$(H$_2$O)$_2$ 
(PM = pyrimidine) \cite{Feyerherm00,Zvyagin04,Wolter03,Wolter05}, and 
KCuGaF$_6$ \cite{Morisaki07,Umegaki09,Umegaki12,Umegaki15}. 
Low-energy properties can be represented by 
the quantum sine-Gordon model. 
Excitations of the solitons and breathers were observed 
in KCuGaF$_6$ \cite{Umegaki15}. 


A nonuniform magnetic field can also be applied to the following spin chains: 
different crystallographic magnetic-ion sites in the spin chains and 
other magnetic-ion sites outside the spin chains.
When magnetic moments on the magnetic-ion sites
outside the spin chains are ordered, 
a different (nonuniform) internal magnetic field is 
applied to the magnetic-ion sites in the spin chains. 


We focus on 
Tb$_3$RuO$_7$, Gd$_3$RuO$_7$, and Nd$_3$RuO$_7$ possessing 
spin-$\frac{3}{2}$ chains formed by the Ru$^{5+}$ ($4d^3$) ions. 
The three compounds are isostructural 
in the high-temperature ($T$) phase. 
The space group is $Cmcm$ 
(No. 63) \cite{Hinatsu14,Bontchev00,Ishizawa08,Groen87,Harada01}.
In Tb$_3$RuO$_7$, Gd$_3$RuO$_7$, and Nd$_3$RuO$_7$, 
a structural phase transition occurs at 
$T_{\rm s} = 402$, 380, and 130 K, respectively, \cite{Wakeshima10} and 
the space groups of the low-$T$ phase are 
$Pna2_1$ (No. 33) \cite{Hinatsu14}, 
$Pna2_1$ (No. 33) \cite{Ida06}, 
and $P2_1/m$ (No. 11) \cite{Harada01}, respectively. 
In the low-$T$ phase,
Tb$_3$RuO$_7$ and Gd$_3$RuO$_7$ are isostructural. 


We describe the arrangements of the Ru sites and 
changes in the Ru and rare-earth ($R$) sites owing to the structural phase transition. 
In the high-$T$ phase, 
the Ru site is unique and uniform Ru chains are formed. 
In each low-$T$ phase, 
two crystallographic Ru (Ru1 and Ru2) sites exist. 
However, the two Ru sites are very similar to each other. 
Figure 1 shows the crystal structure of the low-$T$ phase 
of Tb$_3$RuO$_7$. 
An alternating-bond chain parallel to the $b$ direction 
is formed by the Ru1 and Ru2 ions. 
In the two types of Ru1-Ru2 pairs, 
the Ru1-Ru2 length and Ru1-O-Ru2 angle at room temperature are 
3.67 \AA \ and 141.6$^{\circ}$, and 
3.68 \AA \ and 145.5$^{\circ}$. 
From the angles, 
antiferromagnetic (AF) exchange interactions are expected 
in the Ru1-Ru2 pairs. 
When the Tb magnetic moments are ordered, 
different (nonuniform) internal magnetic fields are 
applied to the Ru1 and Ru2 sites. 
In Nd$_3$RuO$_7$, 
the Ru1 and Ru2 ions form independently uniform chains parallel to the $b$ direction. 
The Ru-Ru length is 3.73 \AA \ in both the chains at 100 K.
The Ru-O-Ru angles are 150.7$^{\circ}$ and 137.6$^{\circ}$ 
in the Ru1-Ru1 and Ru2-Ru2 pairs, respectively. 
When the Nd magnetic moments are ordered, 
the same internal magnetic fields are 
applied to the Ru sites in each chain. 
Regarding 
the change in the $R$ sites due to the structural phase transition, 
the $R1$ and $R2$ sites in the low-$T$ phase 
are the same in the high-$T$ phase. 
The $Ri~(i = 3 - 6)$ sites in the low-$T$ phase 
are the same in the high-$T$ phase. 

\begin{figure}
\begin{center}
\includegraphics[width=16cm]{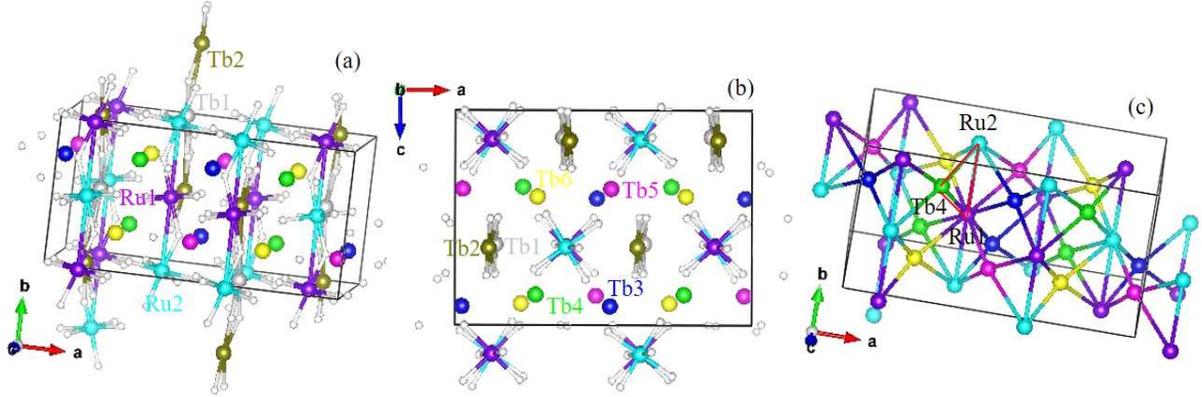}
\caption{
(a)(b)
Crystal structure of the low-$T$ phase of Tb$_3$RuO$_7$, 
drawn using VESTA \cite{Momma11}.
The orthorhombic space group is $Pna2_1$ (No. 33) \cite{Hinatsu14}. 
The lattice constants are
$a=14.588$ \AA, $b=7.345$ \AA, and $c=10.560$ \AA \
at room temperature. 
The rectangle represents a unit cell. 
All the ions are located at the $4a$ sites. 
There are 
two crystallographic Ru sites and 
six crystallographic Tb sites. 
The $y$ coordinates are nearly 0 and 0.5 
at the Ru1, Ru2, Tb1, and Tb2 sites. 
They are nearly 0.25 and 0.75 
at the Tb$i$~($i = 3 - 6$) sites. 
Alternating-bond chains parallel to the $b$ direction are formed 
by the Ru1 and Ru2 ions and 
by the Tb1 and Tb2 ions. 
The crystal structure of the low-$T$ phase of Nd$_3$RuO$_7$ 
is similar to that of Tb$_3$RuO$_7$. 
The following is 
a remarkable difference between the two compounds. 
In Nd$_3$RuO$_7$ with the monoclinic space group $P2_1/m$ (No. 11), 
the Ru1 and Ru2 ions form independently uniform chains parallel to the $b$ direction. 
Similarly, 
the Nd1 and Nd2 ions form independently uniform chains parallel to the $b$ direction. 
(c)
Schematic of 
Ru1, Ru2, and Tb$i$~($i = 3 - 6$) sites. 
Magnetic frustration is expected 
in the Ru1-Tb-Ru2 triangles, such as the red one. 
}
\end{center}
\end{figure}


Next, we describe the magnetic properties of $R_3$RuO$_7$ (R = Tb, Gd, and Nd). 
We can observe a peak indicating an AF transition at $T_{\rm N} = 17$ K and 
a broad maximum at approximately 10 K 
in the specific heat of Tb$_3$RuO$_7$ \cite{Hinatsu14}. 
The peak and maximum are speculated to indicate 
the ordering of the Ru and Tb moments, respectively.
No hysteresis appears between the magnetizations 
measured in the zero-field cooling (ZFC) and field cooling (FC) processes 
in the magnetic field of $\mu_0 H = 0.1$ T. 
In the specific heat of Gd$_3$RuO$_7$, 
two peaks exist at 15 and 9.5 K \cite{Bontchev00,Harada02}. 
It was speculated that 
the Ru and Gd moments were ordered below 15 and 9.5 K, respectively. 
We can see hysteresis between the magnetizations 
measured in the ZFC and FC processes at 0.05 and 0.1 T. 
In Nd$_3$RuO$_7$, 
the specific heat exhibits a peak at 19 K \cite{Harada01}. 
Hysteresis exists between the magnetizations 
measured in the ZFC and FC processes at 0.1 T below 19 K. 
The magnetization curve at 5 K also shows 
hysteresis at $\mu_0 H < 2$ T. 
The hysteresis in the magnetization of Gd$_3$RuO$_7$ and Nd$_3$RuO$_7$ 
indicates the existence of small spontaneous magnetization
(ferromagnetic component). 


Powder neutron-diffraction measurements were performed 
on Nd$_3$RuO$_7$ \cite{Harada01}, and 
the magnetic reflections were observed at 10 K. 
There are two types of propagation vectors 
$\bm{k}_1 = (0, 0, 0)$ and $\bm{k}_2 = (\frac{1}{2}, 0, 0)$ 
for the magnetic structure 
in the standard setting of the crystal structure. 
The $a, b$, and $c$ axes in the published setting \cite{Harada01} 
correspond to the $c, a$, and $b$ axes 
in the standard setting used in this study. 
 Harada {\it et al.} considered that 
the magnetic reflections belonging to $\bm{k}_1$ 
were generated by the Ru ordered moments,
and determined the magnetic structure of the Ru moments. 
The Ru moments are parallel to the $a$ direction and 
are aligned antiferromagnetically in each chain. 
The magnitude of the Ru ordered moments 
was estimated to be $2.2~\mu_{\rm B}$. 
The magnetic structure of the Nd moments 
has not been determined. 


Different (nonuniform) internal magnetic fields are 
applied to the Ru1 and Ru2 sites
when the Tb magnetic moments are ordered in Tb$_3$RuO$_7$. 
An exotic magnetic structure is expected.
The magnetic structure of Nd$_3$RuO$_7$ has not been determined 
but it is useful in understanding 
the magnetic structure of Tb$_3$RuO$_7$. 
Consequently, 
we conducted powder neutron-diffraction experiments 
on Tb$_3$RuO$_7$ and Nd$_3$RuO$_7$. 
The concentration of the magnetic ions is high in $R_3$RuO$_7$. 
If the magnetic entropy change is large, 
$R_3$RuO$_7$ behaves as a magnetic refrigeration material. 
Therefore, 
we investigated the magnetic entropy change in 
Tb$_3$RuO$_7$ and Gd$_3$RuO$_7$, which have 
the large magnetic moments of rare-earth ions. 

\section{Material preparation and experimental methods to study magnetism}

We synthesized the crystalline powders of 
{\it R}$_3$RuO$_7$ ({\it R} = Tb, Gd, and Nd) by using a solid-state reaction. 
The starting materials were 
{\it R}$_2$O$_3$ ({\it R} = Tb, Gd, and Nd), and RuO$_2$ powders. 
The purity was $99.99$~\%. 
A stoichiometric mixture of the powders was sintered in air. 
The sintering temperatures and time were 
1473 K and 24 h for Tb$_3$RuO$_7$ and Gd$_3$RuO$_7$, and 
1523 K and 48 h for Nd$_3$RuO$_7$, respectively. 
We measured the X-ray powder diffraction patterns at room temperature 
by using an X-ray diffractometer (RINT-TTR III, Rigaku). 
For Tb$_3$RuO$_7$ and Gd$_3$RuO$_7$,
the obtained samples were in single phases within experimental accuracy. 
For Nd$_3$RuO$_7$, 
we observed some weak reflections, indicating the existence of other materials 
whose identities could not be determined. 

We measured the magnetization using
the magnetic property measurement system (MPMS) of Quantum Design. 
We carried out neutron-diffraction experiments 
at the Swiss Spallation Neutron Source of
the Paul Scherrer Institut, where 
we used the high-resolution powder diffractometer for thermal neutrons \cite{hrpt}. 
The wavelength of the neutrons ($\lambda$) was 1.886 \AA. 
We carried out 
group theory analyses of the magnetic structures 
using the programs ISODISTORT \cite{Campbell06} and 
BasIreps in the FullProf Suite program package \cite{Rodriguez93}. 
We performed Rietveld refinements of 
the crystal and magnetic structures
using the FullProf Suite program package~\cite{Rodriguez93} 
containing internal tables for 
scattering lengths and magnetic form factors. 

\section{Results}

\subsection{Magnetization}


Figure 2 depicts the $T$ dependence of 
the magnetization [$M(T)$] of 
Tb$_3$RuO$_7$ and Nd$_3$RuO$_7$
measured in a magnetic field of $\mu_0 H = 0.01$ T. 
Hysteresis appears between the magnetizations measured 
in the ZFC and FC processes at low temperatures. 
In Tb$_3$RuO$_7$, 
no hysteresis was reported in 
the magnetization result measured at 0.1 T \cite{Hinatsu14}. 
We were able to observe the hysteresis 
because of the small magnetic field (0.01 T). 
The hysteresis indicates the existence of a tiny spontaneous magnetization. 

\begin{figure}
\begin{center}
\includegraphics[width=8cm]{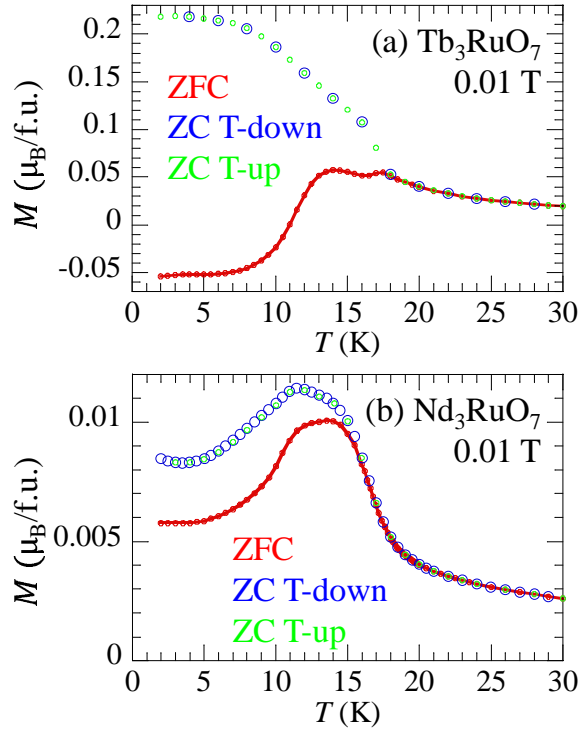}
\caption{
Temperature dependence of the magnetization [$M(T)$] of 
(a) Tb$_3$RuO$_7$ and (b) Nd$_3$RuO$_7$
in a magnetic field of $\mu_0 H = 0.01$ T. 
Red, blue, and green circles represent 
$M(T)$ measured in 
zero-field cooling (ZFC), field cooling (FC), and field warming 
processes, respectively. 
}
\end{center}
\end{figure}


Figure 3 shows the magnetic entropy changes [$-\Delta S_{\rm m}(T)$] 
of Tb$_3$RuO$_7$ and Gd$_3$RuO$_7$
for various strengths of the magnetic field. 
We measured $M(T)$ in the magnetic fields of 0.01 T and 
$0.5 - 5$ T in steps of 0.5 T. 
We derived $-\Delta S_{\rm m}(T)$ 
from the Maxwell relation 
\begin{equation}
-\Delta S_{\rm m}(T) = \int_0^H \left(-\frac{\partial M(T)}{\partial T} \right)_H dH. 
\end{equation}
The red line in each figure indicates $-\Delta S_{\rm m}(T)$ 
for the magnetic field change of 5 T. 
Broad maxima are apparent
 at approximately 19 K and 11 K for 
Tb$_3$RuO$_7$ and Gd$_3$RuO$_7$, respectively. 

\begin{figure}
\begin{center}
\includegraphics[width=8cm]{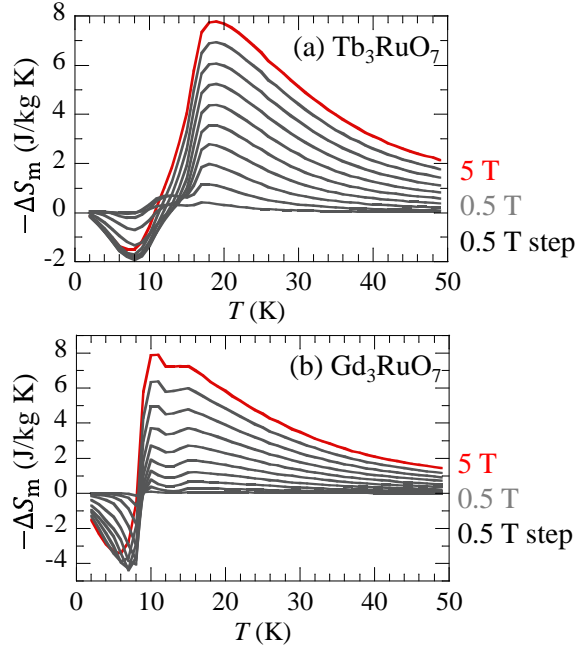}
\caption{
Magnetic entropy change in 
(a) Tb$_3$RuO$_7$ and (b) Gd$_3$RuO$_7$
with change in magnetic field from 0.5 to 5 T in steps of 0.5 T. 
}
\end{center}
\end{figure}

\subsection{Magnetic structure of Tb$_3$RuO$_7$}


Figure 4 depicts the powder neutron-diffraction patterns of Tb$_3$RuO$_7$ 
at 1.5 and 25 K. 
Some reflections are observed only at 1.5 K and 
the other reflections are observed both at 1.5 and 25 K. 
The former reflections appear below 15 K. 
Therefore, these are magnetic reflections. 
Several magnetic reflections are larger or comparable to 
the largest nuclear reflection. 
We can index all the magnetic reflections 
with the propagation vector $\bm{k}_1 = (0, 0, 0)$. 
Table I shows the results of the group theory analysis for $\bm{k}_1$. 
There are four one-dimensional irreducible representations (IRs) 
that appear three times each. 

\begin{figure}
\begin{center}
\includegraphics[width=8cm]{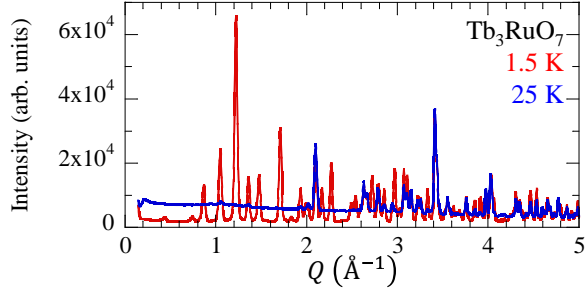}
\caption{
Powder neutron-diffraction patterns of Tb$_3$RuO$_7$
at 1.5 K and 25 K. 
}
\end{center}
\end{figure}

\begin{table*}
\caption{\label{table1}
Group theory analysis for the magnetic structure of Tb$_3$RuO$_7$
calculated using the programs ISODISTORT \cite{Campbell06} and 
BasIreps \cite{Rodriguez93}. 
The character set corresponds to 
the following four symmetry elements \cite{Rodriguez93}: 
Symm(1): 1; 
Symm(2): $2~(0,0,1/2)~0,0,z$; 
Symm(3): $a~(1/2,0,0)~x,1/4,z$; 
Symm(4): $n~(0,1/2,1/2)~1/4,y,z$. 
IR denotes irreducible representation. 
The crystallographic space group is $Pna2_1$ (No. 33). 
The magnetic propagation vector is $\bm{k}_1 = (0, 0, 0)$.
All the ions are located at $4a$ sites. 
The components of the magnetic moments 
are expressed using $u$, $v$, and $w$. 
}
\begin{ruledtabular}
\begin{tabular}{ccccc}
Character set & (1, 1, 1, 1) & (1, 1, -1, -1) & (1, -1, 1, -1) &  (1, -1, -1, 1) \\
IR (ISODISTORT) & $m \Gamma_1$ & $m \Gamma_2$ & $m \Gamma_4$ & $m \Gamma_3$ \\
IR (BasIreps) & IRrep(1) & IRrep(2) & IRrep(3) & IRrep(4) \\
\hline
$(x, y, z)$ & $[u, v, w]$ & $[u, v, w]$ & $[u, v, w]$ & $[u, v, w]$ \\
$(\bar x, \bar y, z+ \frac{1}{2})$ & $[\bar u, \bar v, w]$ & $[\bar u, \bar v, w]$ & $[u, v, \bar w]$ & $[u, v, \bar w]$ \\
$(x+ \frac{1}{2}, \bar y+ \frac{1}{2}, z)$ & $[\bar u, v, \bar w]$ & $[u, \bar v, w]$ & $[\bar u, v, \bar w]$ & $[u, \bar v, w]$ \\
$(\bar x+ \frac{1}{2}, y+ \frac{1}{2}, z+ \frac{1}{2})$ & $[u, \bar v, \bar w]$ & $[\bar u, v, w]$ & $[\bar u, v, w]$ & $[u, \bar v, \bar w]$ \\
\end{tabular}
\end{ruledtabular}
\end{table*}


We determined the magnetic structure of Tb$_3$RuO$_7$ as follows. 
The magnitude of the Tb moment ($g_J J = 9~\mu_{\rm B}$) is 
larger than that of the Ru moment ($g S \sim 3~\mu_{\rm B}$). 
The magnetic reflections are generated mainly by 
the Tb ordered moments 
because the magnitude of the magnetic reflections
is proportional to the square of the magnetic moments. 
Therefore, we first considered only the Tb moments. 
There are eighteen components of the Tb moments 
because of the six crystallographic Tb sites.
It is not realistic to independently refine
the eighteen components. 
In the high-$T$ phase, 
the Tb1 and Tb2 sites are equivalent, and 
the Tb$i$~($i = 3 - 6$) sites are equivalent. 
Therefore, we assumed the constraints that 
the magnitudes of each component of the Tb1 and Tb2 moments 
are identical (|Tb1$j$| = |Tb2$j$|) 
and that 
|Tb3$j$| = |Tb4$j$| = |Tb5$j$| = |Tb6$j$|,
where $j=u, v$, and $w$. 
We carried out Rietveld refinements 
using only the Tb moments for the pattern at 1.5 K.  
We found that the best-fit candidate was $m \Gamma_3$ 
and that 
the Tb1$u$, Tb1$w$, and Tb3$v$ components  
were negligible. 


Next, we performed the Rietveld refinements 
using the Ru moments as well as the Tb moments 
for the pattern at 1.5 K. 
We evaluated that 
the $v$ and $w$ components of the Ru moments were finite
and that 
the $u$ components were negligible. 
The $u$ components are ferromagnetic in $m \Gamma_3$.
Therefore, the evaluation is consistent with 
the tiny spontaneous magnetization seen in Fig. 2. 
Figure 5 shows the $T$ dependence of 
the magnitude of the Ru1 and Ru2 moments.  
It seems unphysical that 
the magnitude of the Ru2 moment at 15 K 
is larger than that at 1.5 K. 

\begin{figure}
\begin{center}
\includegraphics[width=8cm]{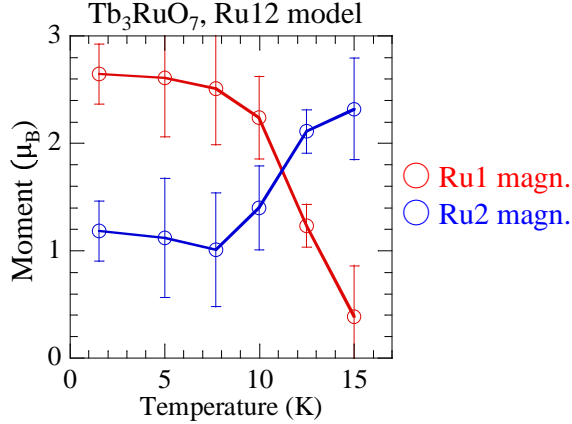}
\caption{
Temperature dependence of 
the magnitudes of the Ru1 and Ru2 moments 
in Tb$_3$RuO$_7$
evaluated in Rietveld refinements 
using both the Ru1 and Ru2 moments (Ru12 model). 
}
\end{center}
\end{figure}


We considered other models with ordering of
only the Ru1 moments (Ru1 model) and 
only the Ru2 moments (Ru2 model). 
The blue line in Fig. 6(a) 
indicates the result of Rietveld refinements 
using the Ru1 model, 
and can explain the observed diffraction pattern at 1.5 K (red circles). 
In contrast, the Ru2 model 
cannot explain the observed pattern well. 
The blue and green lines in Fig. 6(b) 
indicate the results of the Rietveld refinements 
using the Ru1 and Ru2 models, respectively. 
As is seen at around $Q = 0.43, 0.74$, and 0.96~\AA$^{-1}$, 
the consistency between the observed and refined patterns 
is better in the Ru1 model than in the Ru2 model. 

\begin{figure}
\begin{center}
\includegraphics[width=8cm]{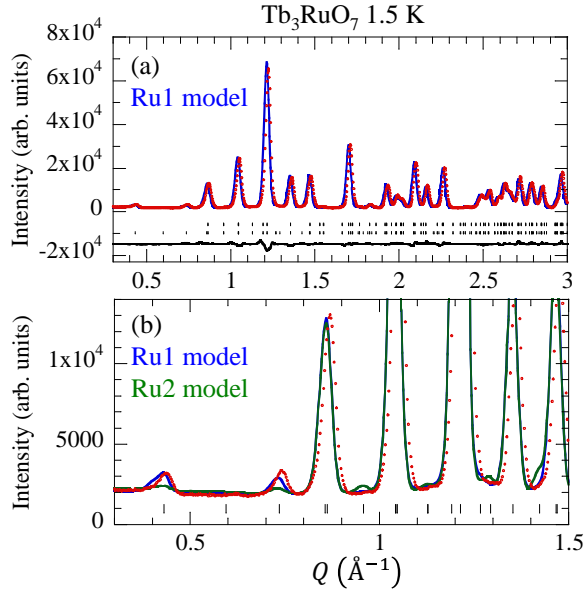}
\caption{
Powder neutron-diffraction patterns (circles) at 1.5 K 
of Tb$_3$RuO$_7$. 
(a)
The line on the measured pattern 
portrays the Rietveld-refined pattern obtained using 
the Ru1 model. 
We used atomic parameters 
determined by 
Rietveld refinements for the pattern at 25 K 
using $Pna2_1$. 
The line at the bottom portrays the difference 
between the measured and Rietveld-refined patterns.  
The upper and lower hash marks represent 
the positions of 
the nuclear and magnetic reflections, respectively.
The reliability indexes of the refinement are 
$R_{\rm p}=3.12$~\%, $R_{\rm wp}=4.13$~\%, and $R_{\rm exp}=0.77$~\%. 
(b)
The blue and green lines on the measured pattern 
portray the Rietveld-refined pattern obtained using 
the Ru1 and Ru2 models, respectively. 
The hash marks represent 
the positions of the magnetic reflections.
}
\end{center}
\end{figure}


Figure 7 shows the magnetic structure of Tb$_3$RuO$_7$ at 1.5 K. 
The Ru1 and Ru2 moments are ordered and disordered (paramagnetic), respectively, 
giving an appearance of a partially disordered (PD) state. 
The values of $|v|$ and $|w|$ of the Ru1 moments 
are 2.38(3) and $2.18(4)~\mu_{\rm B}$, respectively. 
The magnitude is $3.23(5)~\mu_{\rm B}$ and corresponds to $g = 2.15$. 
The order of the Tb1 and Tb2 moments is AF 
in each Tb1-Tb2 chain parallel to the $b$ direction. 
The Tb1 and Tb2 ordered moments are parallel to the $b$ direction and 
their magnitude is $8.58(2)~\mu_{\rm B}$. 
The Tb$i$~($i = 3 - 6$) moments form a noncollinear magnetic structure.  
The values of $|u|$ and $|w|$ are 
$3.70(1)$ and $7.72(2)~\mu_{\rm B}$, respectively.  
The magnitude is $8.56(2)~\mu_{\rm B}$. 
The magnitude of all the Tb moments is slightly smaller than
the theoretical value ($g_J J = 9~\mu_{\rm B}$). 

\begin{figure}
\begin{center}
\includegraphics[width=16cm]{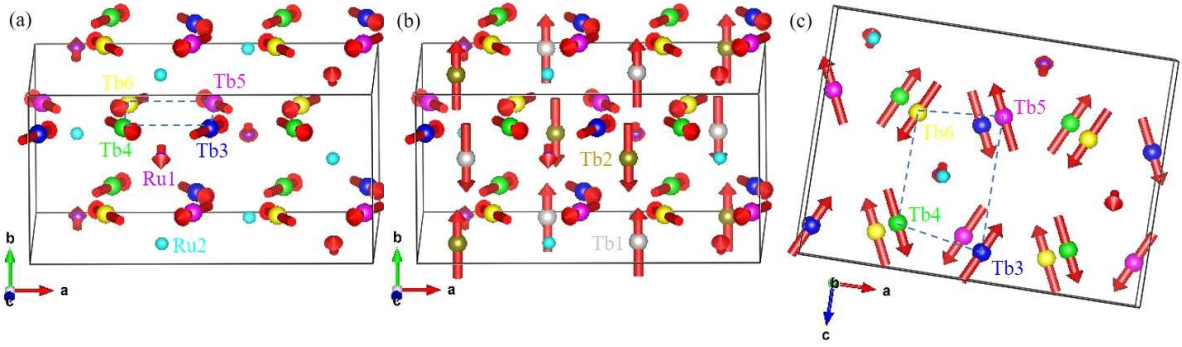}
\caption{
Magnetic structure of Tb$_3$RuO$_7$ at 1.5 K, 
drawn using VESTA \cite{Momma11}. 
The IR is $m \Gamma_3$. 
The rectangle represents a unit cell.
Tb1 and Tb2 are omitted in (a) and (c). 
The dashed lines in (a) and (c) represent a four-Tb ring, 
where the chiral vector is defined. 
}
\end{center}
\end{figure}


Figure 8 depicts the $T$ dependence of 
the components and magnitudes of 
the ordered magnetic moments. 
Each component is almost constant below 10 K and 
decreases with increase in $T$ above 10 K. 
The $v$ components decrease more rapidly than the other components. 
We speculated that 
the rotations of the Tb1 and Tb2 moments occurred 
at approximately 12.5 K. 
In contrast, the rotation of the Ru1 moment must be tiny 
because the $v$ and $w$ components are already considered 
and because the ferromagnetic $u$ component must be negligible. 
We carried out Rietveld refinements 
on the diffraction pattern at 12.5 K 
by adding the Tb1$u$, Tb1$w$, Tb2$u$, and Tb2$w$ components 
with the constraint that Tb1$u$ = -Tb2$u$
because of the tiny spontaneous magnetization. 
We evaluated that 
Tb1$u$ = -Tb2$u$ = 0.67(19), 
Tb1$w$ = 0.09(16), and Tb2$w$=0.45(17)
in the unit of $\mu_{\rm B}$. 
The values of the other components 
were almost unchanged by adding these components. 
The magnitude of the Tb1$v$ and Tb2$v$ components 
is $4.36(5) \mu_{\rm B}$ and 
is much larger than 
that of the Tb1$u$, Tb1$w$, Tb2$u$, and Tb2$w$ components. 
Therefore, 
any rotation of the Tb1 and Tb2 moments would be small.  
Next, we discuss the above specific-heat results \cite{Hinatsu14}. 
The peak at $T_{\rm N}=17$ K and broad maximum at approximately 10 K 
in the specific heat
were speculated to indicate 
the ordering of the Ru and Tb moments, respectively.
We obtained the result that
both the Ru1 and Tb moments were ordered at 15 K.
No transition was observed between 15 K and 17 K. 
Therefore, 
both the Ru1 and Tb moments are ordered simultaneously 
at $T_{\rm N}=17$ K. 
The relatively rapid changes in the $v$ components probably 
generated 
the broad maximum at approximately 10 K.  

\begin{figure}
\begin{center}
\includegraphics[width=8cm]{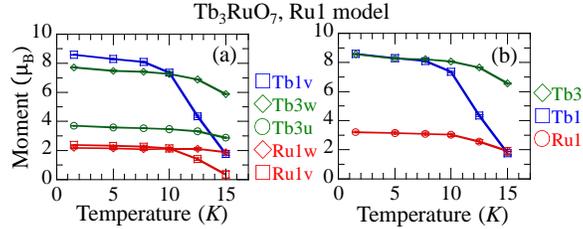}
\caption{
(a)
Temperature dependence of the components of 
the Ru1, Tb1, and Tb3 moments in Tb$_3$RuO$_7$,
evaluated in Rietveld refinements 
using the Ru1 model. 
In the Rietveld refinements, 
we used the constraints
that |Tb1$v$| = |Tb2$v$| 
and 
|Tb$ij$|~($i = 3 - 6$, $j=u$ and $w$) are identical. 
Although error bars are shown, 
they are not clearly visible in this scale of the moment. 
(b) 
Temperature dependence of the magnitudes of 
the Ru1, Tb1, and Tb3 moments. 
}
\end{center}
\end{figure}

\subsection{Magnetic structure of Nd$_3$RuO$_7$}


Figure 9(a) depicts the powder neutron-diffraction patterns of Nd$_3$RuO$_7$ 
at 1.5 and 25 K. 
Some reflections are observed only at 1.5 K and 
the other reflections are observed both at 1.5 and 25 K. 
The former reflections appear below 17.5 K. 
Therefore, these are magnetic reflections. 
Our measurements confirmed the existence of two types of propagation vectors 
$\bm{k}_1 = (0, 0, 0)$ and $\bm{k}_2 = (\frac{1}{2}, 0, 0)$, 
as has been reported by Harada et al. \cite{Harada01}. 
The results of the group theory analysis calculated by 
the programs ISODISTORT \cite{Campbell06} and
BasIreps \cite{Rodriguez93}
are summarized in Table II. 
In comparison with Tb$_3$RuO$_7$ 
with only one magnetic propagation vector ($\bm{k}_1$) and 
all magnetic ions at the same crystallographic site ($4a$), 
the situation in Nd$_3$RuO$_7$ is more complicated. 
There are two magnetic propagation vectors, and 
the magnetic ions are distributed across five different crystallographic sites. 

\begin{figure}
\begin{center}
\includegraphics[width=8cm]{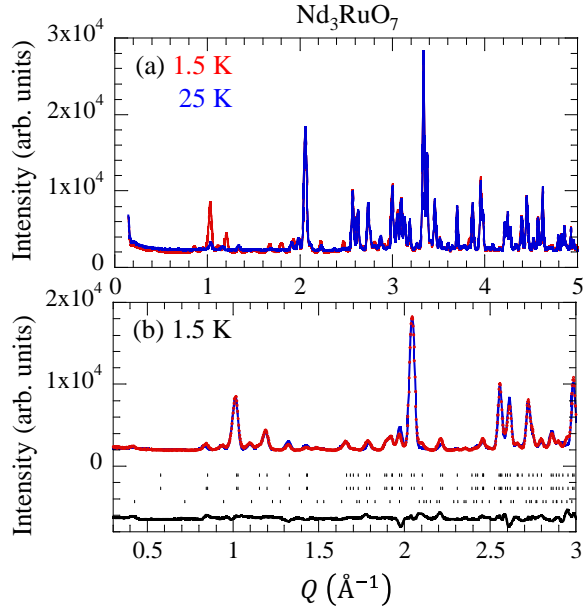}
\caption{
(a)
Powder neutron-diffraction patterns of Nd$_3$RuO$_7$
at 1.5 K and 25 K. 
(b)
The results of Rietveld refinements 
for the pattern at 1.5 K (circles). 
We applied the atomic parameters 
determined by 
Rietveld refinements for the pattern at 25 K 
using $P2_1/m$.  
The line on the measured pattern 
portrays the Rietveld-refined pattern. 
The line at the bottom portrays the difference 
between the measured and Rietveld-refined patterns.  
The upper, middle, and lower hash marks represent 
the positions of 
nuclear reflections, magnetic reflections of $\bm{k}_1$, and 
magnetic reflections of $\bm{k}_2$. 
The reliability indexes of the refinement are 
$R_{\rm p}=5.01$~\%, $R_{\rm wp}=6.70$~\%, and $R_{\rm exp}=0.96$~\%. 
}
\end{center}
\end{figure}

\begin{table*}
\caption{\label{table2}
Group theory analysis for the magnetic structure of Nd$_3$RuO$_7$
calculated by the programs ISODISTORT \cite{Campbell06} and 
BasIreps \cite{Rodriguez93}. 
The character set corresponds to 
the following four symmetry elements \cite{Rodriguez93}:
Symm(1): 1; 
Symm(2): $2~(0,1/2,0)~0,y,0$; 
Symm(3): $\bar 1~0,0,0$; 
Symm(4): $m~x,1/4,z$.
IR denotes irreducible representation. 
The crystallographic space group is $P2_1/m$ (No. 11). 
The magnetic propagation vectors are 
$\bm{k}_1 = (0, 0, 0)$ and $\bm{k}_2 = (\frac{1}{2}, 0, 0)$. 
The components of the magnetic moments 
are expressed as $u$, $v$, and $w$. 
The blank cells indicate that the 
magnetic order of the moments on the corresponding site is impossible.
}
\begin{ruledtabular}
\begin{tabular}{lll|cc|cc|cc|cc}
\multicolumn{3}{c|}{Character set} & \multicolumn{2}{c|}{(1, 1, 1, 1)} & \multicolumn{2}{c|}{(1, 1, -1, -1)} & \multicolumn{2}{c|}{(1, -1, 1, -1)} & \multicolumn{2}{c}{(1, -1, -1, 1)} \\
\multicolumn{3}{c|}{Propagation vector} & $\bm{k}_1$ & $\bm{k}_2$ & $\bm{k}_1$ & $\bm{k}_2$ & $\bm{k}_1$ & $\bm{k}_2$ & $\bm{k}_1$ & $\bm{k}_2$ \\
\multicolumn{3}{c|}{IR (ISODISTORT)} & $m \Gamma_1^+$ & $mY_1^+$ & $m \Gamma_1^-$ & $mY_1^-$ & $m \Gamma_2^+$ & $mY_2^+$ & $m \Gamma_2^-$ & $mY_2^-$ \\
\multicolumn{3}{c|}{IR (BasIreps)} & \multicolumn{2}{c|}{IRrep(1)} & \multicolumn{2}{c|}{IRrep(2)} & \multicolumn{2}{c|}{IRrep(3)} & \multicolumn{2}{c}{IRrep(4)} \\
\hline
Ru1 & $2c$ & $(0, 0, \frac{1}{2})$ & $[u, v, w]$ & $[u, v, w]$ & & & $[u, v, w]$ & $[u, v, w]$ & & \\
& & $(0, \frac{1}{2}, \frac{1}{2})$ & $[\bar u, v, \bar w]$ & $[\bar u, v, \bar w]$ & & & $[u, \bar v, w]$ & $[u, \bar v, w]$ & & \\
\hline
Ru2 & $2b$ & $(\frac{1}{2}, 0, 0)$ & $[u, v, w]$ & & & $[u, v, w]$ & $[u, v, w]$ & & & $[u, v, w]$ \\
& & $(\frac{1}{2}, \frac{1}{2}, 0)$ & $[\bar u, v, \bar w]$ & & & $[u, \bar v, w]$ & $[u, \bar v, w]$ & & & $[\bar u, v, \bar w]$ \\
\hline
Nd1 & $2d$ & $(\frac{1}{2}, 0, \frac{1}{2})$ & $[u, v, w]$ & & & $[u, v, w]$ & $[u, v, w]$ & & & $[u, v, w]$ \\
& & $(\frac{1}{2}, \frac{1}{2}, \frac{1}{2})$ & $[\bar u, v, \bar w]$ & & & $[u, \bar v, w]$ & $[u, \bar v, w]$ & & & $[\bar u, v, \bar w]$ \\
\hline
Nd2 & $2a$ & $(0, 0, 0)$ & $[u, v, w]$ & $[u, v, w]$ & & & $[u, v, w]$ & $[u, v, w]$ & & \\
& & $(0, \frac{1}{2}, 0)$ & $[\bar u, v, \bar w]$ & $[\bar u, v, \bar w]$ & & & $[u, \bar v, w]$ & $[u, \bar v, w]$ & & \\
\hline
Nd3-Nd6 & $2e$ & $(x, \frac{1}{4}, z)$ & $[0, v, 0]$ & $[0, v, 0]$ & $[u, 0, w]$ & $[u, 0, w]$ & $[u, 0, w]$ & $[u, 0, w]$ & $[0, v, 0]$ & $[0, v, 0]$ \\
& & $(\bar x, \frac{3}{4}, \bar z)$ & $[0, v, 0]$ & $[0, v, 0]$ & $[\bar u, 0, \bar w]$ & $[\bar u, 0, \bar w]$ & $[u, 0, w]$ & $[u, 0, w]$ & $[0, \bar v, 0]$ & $[0, \bar v, 0]$ \\
\end{tabular}
\end{ruledtabular}
\end{table*}


We determined the magnetic structure of Nd$_3$RuO$_7$ as follows. 
As is seen in Fig. 9, 
there are several strong magnetic reflections for $\bm{k}_1$, 
whereas all the magnetic reflections for $\bm{k}_2$ are weak. 
The difference indicates that 
most magnetic moments generate the magnetic reflections of $\bm{k}_1$. 
After the initial Rietveld refinements, 
we found that 
the best-fit candidates for $\bm{k}_1$ and $\bm{k}_2$ were 
$m \Gamma_1^+$ and $m Y_1^+$, respectively, 
and that 
only the Ru1 moments obey $m Y_1^+$.  
Note that 
both $m \Gamma_1^+$ and $m Y_1^+$ 
belong to the same character set, as shown in Table II. 

It is not realistic to independently refine
all the components of the magnetic moments. 
In the high-$T$ phase, 
the Nd1 and Nd2 sites are equivalent, and the
Nd$i$~($i = 3 - 6$) sites are equivalent. 
Therefore, we assumed the constraints 
that |Nd1$j$| = |Nd2$j$| ($j=u, v$, and $w$) and 
|Nd$iv$|~($i = 3 - 6$) are identical. 
Nd$iu$ and Nd$iw$~($i = 3 - 6$) are zero because of symmetry.  
In $m \Gamma_1^+$, the $v$ components are ferromagnetic and 
the source of the tiny spontaneous magnetization in $M(T)$. 
Thus, we assumed the additional constraints 
that Nd1$v$ = -Nd2$v$, 
the sum of Nd$iv$~($i = 3 - 6$) = 0, and Ru2$v$ = 0. 
We conducted Rietveld refinements 
to determine the magnetic structure.  
The blue line in Fig. 9(b) 
indicates the results of the Rietveld refinements 
and can explain the observed diffraction pattern at 1.5 K (red circles). 


Figure 10 shows the magnetic structure of Nd$_3$RuO$_7$ at 1.5 K.  
It is significantly different from the magnetic structure of 
Tb$_3$RuO$_7$ at 1.5 K despite the similar crystal structures. 
The values of $|u|$ and $|w|$ of the Ru1 moments 
are 1.84(40) and $1.93(30)~\mu_{\rm B}$, respectively, and 
those of the Ru2 moments 
are 1.98(18) and $2.06(11)~\mu_{\rm B}$, respectively. 
The values of Ru1$v$ and Ru2$v$ are negligible. 
The magnitudes of the Ru1 and Ru2 moments are 
2.67(50) and $2.86(21)~\mu_{\rm B}$, respectively, and 
is slightly smaller than the classical value ($\sim 3~\mu_{\rm B}$), 
suggesting the existence of quantum fluctuation 
due to the one dimensionality. 
In contrast to Tb$_3$RuO$_7$, no PD state of the Ru moments appears. 
The order of the Ru moments is AF in each chain parallel to the $b$ direction. 
The AF alignment is consistent with 
the signs of the intrachain exchange interactions 
expected from the Ru-O-Ru angles 
(150.7$^{\circ}$ and 137.6$^{\circ}$). 
The magnetic structures of the Ru1 and Ru2 moments indicate that 
the interchain exchange interactions along the $a$ direction are 
antiferromagnetic and ferromagnetic, respectively. 
The present magnetic structure is different from 
the previously reported structure \cite{Harada01}. 
We performed Rietveld refinements for 
the magnetic reflections belonging to $\bm{k}_1$ 
by using moments other than the Ru1 moments,  
whereas
Harada {\it et al.} performed Rietveld refinements 
using the Ru1 and Ru2 moments. 
Therefore, our results are different from those reported in their study. 


The values of $|u|$ and $|w|$ of the Nd1 and Nd2 moments 
are 1.81(14) and $1.99(8)~\mu_{\rm B}$, respectively, 
whereas that of $|v|$ is negligible. 
The magnitude is $2.68(16)~\mu_{\rm B}$. 
The order of the Nd1 and Nd2 moments is AF in each chain parallel to the $b$ direction. 
The magnitude of the Nd$i$~($i = 3 - 6$) moments parallel to the $b$ direction 
is $2.01(6)~\mu_{\rm B}$.
The Nd$i$~($i = 3 - 6$) moments form a collinear magnetic structure 
that is significantly different from 
the noncollinear magnetic structure of 
the Tb$i$~($i = 3 - 6$) moments in Tb$_3$RuO$_7$. 
The theoretical value of the Nd moment ($g_J J = 3.27~\mu_{\rm B}$) 
is slightly larger than those of the Nd1 and Nd2 moments, and 
is much larger than those of the Nd$i$~($i = 3 - 6$) moments. 

\begin{figure}
\begin{center}
\includegraphics[width=16cm]{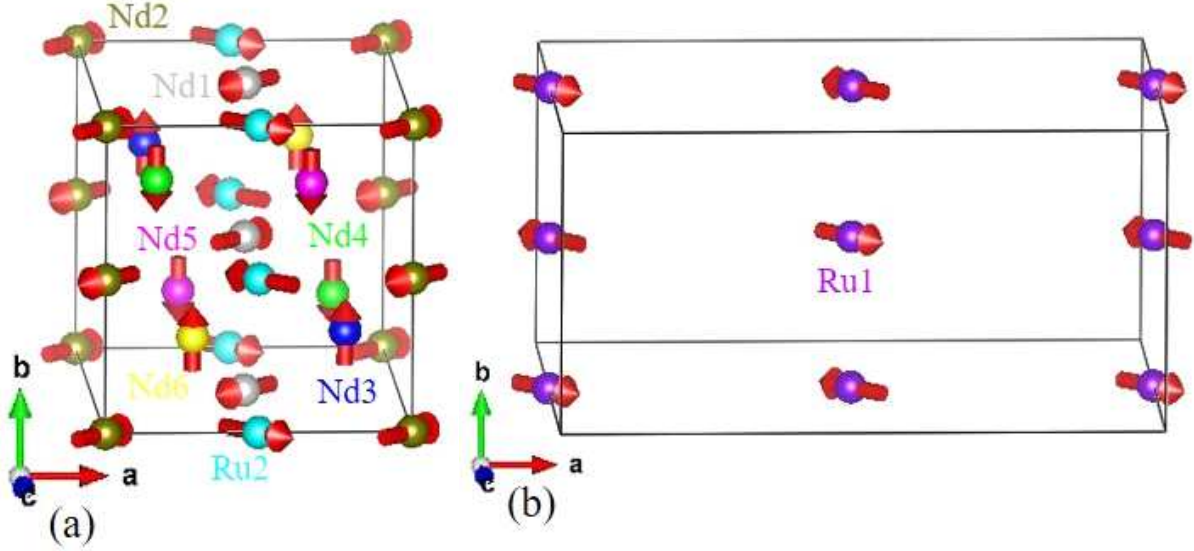}
\caption{
Magnetic structure of Nd$_3$RuO$_7$ at 1.5 K, 
drawn using VESTA \cite{Momma11}. 
The IR is 
(a) $m \Gamma_1^+$ for $\bm{k}_1 = (0, 0, 0)$ and 
(b) $mY_1^+$ for $\bm{k}_2 = (\frac{1}{2}, 0, 0)$. 
The rectangle represents the magnetic unit cell.
}
\end{center}
\end{figure}


Figure 11 depicts the $T$ dependence of 
the components and magnitudes of 
the ordered magnetic moments. 
The magnitude of the Nd moments 
decrease to low values at 17.5 K, whereas 
that of the Ru moments remains large at 17.5 K. 
As $T$ is lowered from $T_{\rm N} = 19$ K, 
the Ru and Nd moments appear to increase rapidly and gradually, respectively. 

\begin{figure}
\begin{center}
\includegraphics[width=8cm]{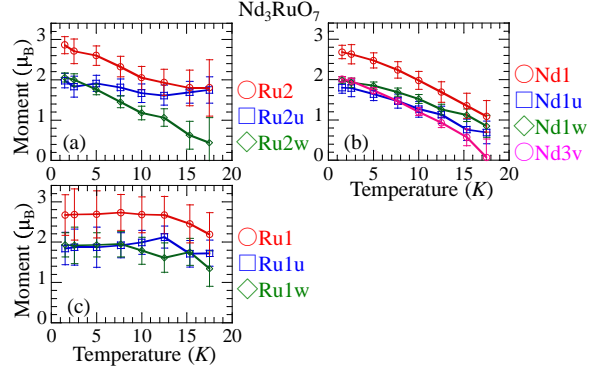}
\caption{
Temperature dependence of 
the components and magnitude of magnetic moments 
in Nd$_3$RuO$_7$, evaluated through Rietveld refinements; 
(a) Ru2, (b) Nd1 and Nd3, and (c) Ru1. 
In the Rietveld refinements, 
we used the constraints
|Nd1$j$| = |Nd2$j$|, where $j=u$ and $w$, and 
|Nd$iv$|~($i = 3 - 6$) are identical. 
}
\end{center}
\end{figure}

\section{Discussion}

\subsection{Magnetic entropy change}

We compared the magnetic entropy changes of 
Tb$_3$RuO$_7$ and Gd$_3$RuO$_7$ with 
those of other oxides containing rare-earth ions. 
Figure 12 shows the maximum magnetic entropy change 
($-\Delta S_{\rm m, max}$) versus 
temperature ($T_{\rm max}$) at which 
the magnetic entropy change is the maximum. 
The change in the magnetic field is 5 T. 
As a rough trend, 
$-\Delta S_{\rm m, max}$ increases with decreasing  $T_{\rm max}$ \cite{Franco18}. 
In Tb$_3$RuO$_7$ and Gd$_3$RuO$_7$, 
the values of $-\Delta S_{\rm m, max}$ are not very large 
for the values of $T_{\rm max}$ 
because  
the magnetic orders are not ferromagnetic 
in the two compounds.

\begin{figure}
\begin{center}
\includegraphics[width=8cm]{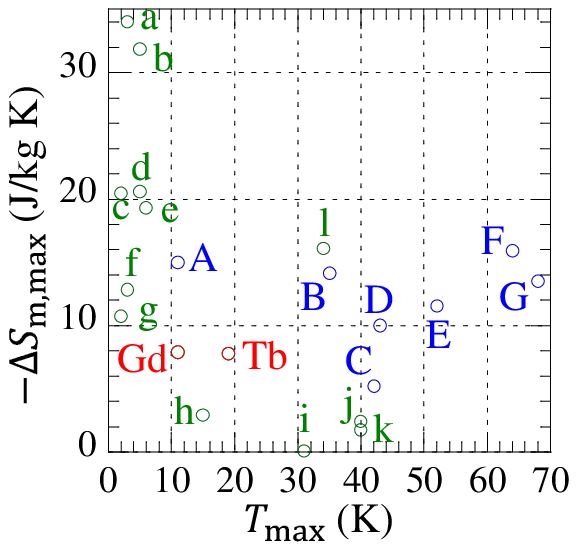}
\caption{
$-\Delta S_{\rm m, max}$ versus $T_{\rm max}$ of 
several oxides. 
Tb: Tb$_3$RuO$_7$, Gd: Gd$_3$RuO$_7$, 
a: GdFeO$_3$ \cite{Das17}, b: Ho$_2$O$_3$ \cite{Boutahar17}, 
c: Er$_6$WO$_{12}$ \cite{Hase21}, d: Ho$_6$WO$_{12}$ \cite{Hase21}, 
e: Dy$_6$WO$_{12}$ \cite{Hase21}, f: Gd$_2$O$_3$ \cite{Paul16}, 
g: Sr$_2$DyNbO$_6$ \cite{Hase20}, h: Tm$_6$MoO$_{12}$ \cite{Hase21}, 
i: Sr$_2$YRuO$_6$ \cite{Hase20}, j: Sr$_2$DyRuO$_6$ \cite{Hase20}, 
k: Sr$_2$TbRuO$_6$ \cite{Hase20}, l: Er$_2$Mn$_2$O$_7$ \cite{Cai17}, 
A: TbMnO$_3$ ($H // a$) \cite{Jin11}, B: GdTiO$_3$ ($H // c$) \cite{Omote19}, 
C: YbTiO$_3$ ($H // c$) \cite{Su13}, D: ErTiO$_3$ ($H // c$) \cite{Su13},
E: HoTiO$_3$ ($H // c$) \cite{Su11}, F: DyTiO$_3$ ($H // c$) \cite{Su12a}, 
G: TmTiO$_3$ ($H // c$) \cite{Su12b}. 
Lowercase and uppercase letters indicate 
the data for powders and single crystals, respectively. 
}
\end{center}
\end{figure}

\subsection{Origin of the partially disordered state}

We consider the origin of the PD state in Tb$_3$RuO$_7$. 
As shown in Table III, 
several antiferromagnets possessing magnetic frustration in exchange interactions
constitute the PD state. 
The materials belonging to group A 
have multiple crystallographic magnetic-ion sites. 
In the spin-$\frac{1}{2}$ frustrated three-leg-ladder Heisenberg antiferromagnets 
Cu$_3$(OH)$_4${\it A}O$_4$ ({\it A} = S or Se),  
the Cu moments within the two outer legs are ordered, whereas
the Cu moments within the inner leg remain random \cite{Vilminot03,Vilminot07}. 
In the spin-$\frac{1}{2}$ frustrated quasi-one-dimensional antiferromagnet 
Cu$_3$Mo$_2$O$_9$, 
the Cu moments within the dimers are ordered, whereas
the Cu moments within the chains remain random \cite{Hase15}.  
The materials belonging to group C 
show a PD state
in spite of a unique crystallographic magnetic-ion site. 

\begin{table*}
\caption{\label{table3}
Antiferromagnets showing a partially disordered (PD) state. 
The materials belonging to the groups A and B
have multiple crystallographic magnetic-ion sites. 
The sites are 
clearly different from each other in group A, and 
are very similar to each other in group B. 
The materials belonging to group C 
have unique crystallographic magnetic-ion site. 
}
\begin{ruledtabular}
\begin{tabular}{ccc}
group A & group B & group C \\
\hline
Cu$_3$(OH)$_4$SO$_4$ \cite{Vilminot03,Vilminot07} & Tb$_3$RuO$_7$ & CsCoCl$_3$ \cite{Mekata76} \\
Cu$_3$(OH)$_4$SeO$_4$ \cite{Vilminot07}  &  & CuFeO$_2$ \cite{Mekata93} \\
Cu$_3$Mo$_2$O$_9$ \cite{Hase15} &  & Ag$_2$CrO$_2$ \cite{Matsuda12} \\
 &  & Gd$_2$Ti$_2$O$_7$ \cite{Stewart04} \\
 &  & GeNi$_2$O$_4$ \cite{Matsuda08} \\
\end{tabular}
\end{ruledtabular}
\end{table*}

Magnetic frustration in exchange interactions 
can exist in Tb$_3$RuO$_7$. 
For example, 
let us consider the Ru1-Tb4-Ru2 triangle indicated in Fig. 1(c). 
There are eight similar Ru1-Tb-Ru2 triangles per Ru site. 
The Ru1-Ru2 interactions are probably AF 
based on the Ru1-O-Ru2 angles (141.6$^{\circ}$ and 145.5$^{\circ}$).  
The interactions between the Ru and Tb moments must exist 
because both the Ru1 and Tb moments are ordered 
at the same $T$ (17 K). 
Magnetic frustration in the Ru1-Tb-Ru2 triangles is possible 
when the signs of the Ru1-Tb and Ru2-Tb interactions are the same. 
In contrast, it appears that
the Ru-Ru interactions alone cannot generate magnetic frustration
because of the following facts.  
The Ru1-Ru2 chains are almost linear and 
are well separated from one another. 
The next-nearest-neighbor Ru-Ru interaction in the chains and 
interchain interactions may be very weak. 


We speculate that 
a chiral vector generated by four Tb moments 
may also be the origin of the magnetic frustration. 
The chiral vector is defined as 
\begin{equation}
\bm{V}_{\rm c} = \bm{M}_1 \times \bm{M}_2 + \bm{M}_2 \times \bm{M}_3 +
\bm{M}_3 \times \bm{M}_4 + \bm{M}_4 \times \bm{M}_1,
\end{equation}
where $\bm{M}_i \ (i = 1 - 4)$ denote the magnetic moment vectors. 
For example, in the Tb4356 ring, indicated by the dashed lines in Fig. 7, 
we let the Tb4, Tb3, Tb5, and Tb6 moments correspond to
$\bm{M}_1$, $\bm{M}_2$, $\bm{M}_3$, and $\bm{M}_4$, respectively.
The chiral vector is calculated as (0, 228, 0) $\mu_{\rm B}^2$ 
in the Tb4356 ring located at approximately $y \sim 0.75$. 
The directions of the Tb moment vectors 
in the Tb6534 ring located at approximately $y \sim 0.25$ 
are opposite to those of the respective Tb moment vectors 
in the Tb4356 ring connected by the dashed lines. 
The chiral vector is (0, 228, 0) $\mu_{\rm B}^2$ 
in the Tb6534 ring as well. 
Consequently, the chiral vectors are parallel (ferromagnetic)
along the $b$ direction.  
The Ru1-Ru2 chains penetrate the rings formed by the four Tb sites.  
As described, AF exchange interactions are expected 
between the nearest-neighbor Ru1 and Ru2 spins in the chains. 
If the interaction between the chiral vector and Ru spin exists, 
magnetic frustration may occur. 
In contrast, in Nd$_3$RuO$_7$, the Nd$i$~($i = 3 - 6$) moments form 
a collinear magnetic structure. 
Therefore, no chiral vector exists. 


We consider that 
the main source of the PD state in Tb$_3$RuO$_7$ is 
magnetic frustration in exchange interactions. 
However, all the Ru moments are ordered in Nd$_3$RuO$_7$ 
although a similar magnetic frustration can be expected. 
Considering materials with a unique crystallographic magnetic-ion site
(group C), 
the absence of a PD state in Nd$_3$RuO$_7$ 
cannot be explained by the fact that the Ru sites in each chain are equivalent.   
The magnitude of the Nd moment ($g_J J = 3.27~\mu_{\rm B}$) is 
smaller than that of the Tb moment ($9~\mu_{\rm B}$). 
Therefore, the magnetic frustration may be weaker in Nd$_3$RuO$_7$ 
than in Tb$_3$RuO$_7$, leading to
the absence of a PD state in Nd$_3$RuO$_7$.  
There is another difference between the two compounds. 
The internal magnetic field generated by the $R$ ordered moments at the Ru1 sites 
is different from that at the Ru2 sites in Tb$_3$RuO$_7$, 
whereas the magnetic field is the same at the Ru sites in each chain 
of Nd$_3$RuO$_7$. 
We speculate that 
the different (nonuniform) magnetic field increases 
the difference in the properties between the
Ru1 and Ru2 moments (ordered and disordered, respectively).  

\section{Summary}

We performed powder neutron-diffraction experiments on 
Tb$_3$RuO$_7$ and Nd$_3$RuO$_7$
to determine the magnetic structures. 
There are two crystallographic Ru sites and 
six crystallographic {\it R} ({\it R} = Tb or Nd) sites 
in the low-$T$ phase. 
In Tb$_3$RuO$_7$, 
alternating-bond spin-$\frac{3}{2}$ Ru1-Ru2 chains 
are formed parallel to the $b$ direction. 
The Ru and Ru2 moments are 
ordered and disordered (paramagnetic), respectively, below 15 K. 
This result indicates the appearance of a PD state, 
although the two Ru sites are very similar to each other.   
The order of the Tb moments is AF 
in each Tb1-Tb2 chain parallel to the $b$ direction. 
The Tb$i$~($i = 3 - 6$) moments 
form a noncollinear magnetic structure. 
In Nd$_3$RuO$_7$, 
the Ru1 and Ru2 ions form independently uniform chains parallel to the $b$ direction. 
The order of the Ru moments is AF in each chain. 
The order of the Nd1 and Nd2 moments is also AF 
in each chain parallel to the $b$ direction. 
The Nd$i$~($i = 3 - 6$) moments 
form a collinear magnetic structure. 
The main cause of the PD state in Tb$_3$RuO$_7$ is 
probably the magnetic frustration in the exchange interactions. 
Based on the difference in the magnetic structures 
between Tb$_3$RuO$_7$ and Nd$_3$RuO$_7$, 
we speculate that 
the chiral vector formed by the Tb3, Tb4, Tb5, and Tb6 moments 
may also generate magnetic frustration. 
The internal magnetic field generated by the Tb ordered moments at the Ru1 sites 
is different from that at the Ru2 sites.
We speculate that 
the different (nonuniform) internal magnetic field increases 
the difference in the properties between 
the Ru1 and Ru2 moments.  


We investigated 
the temperature dependence of the magnetic entropy changes of 
Tb$_3$RuO$_7$ and Gd$_3$RuO$_7$.  
When the magnetic-field change is 5 T,  
broad maxima are observed 
at approximately $T_{\rm max} = 19$ K and 11 K, and  
the maximum magnetic entropy changes ($-\Delta S_{\rm m, max}$) are 
7.78 and 7.92 J/(kg K) 
for Tb$_3$RuO$_7$ and Gd$_3$RuO$_7$, respectively. 
The values of $-\Delta S_{\rm m, max}$ are not very large 
for the values of $T_{\rm max}$ 
because  
the magnetic orders are not ferromagnetic 
in the two compounds.

\begin{acknowledgments}

This work was supported by the 
Japan Society for the Promotion of Science (JSPS) 
KAKENHI Grant Number 18K03551, 
a grant for 
advanced measurement and characterization technologies 
accelerating materials innovation (PF2050)
from National Institute for Materials Science (NIMS), and 
JST-Mirai Program Grant Number JPMJMI18A3, Japan.
We are grateful 
to Takashi Mochiku, Hiroaki Mamiya, Masamichi Nishino, 
Noriki Terada, and Naohito Tsujii at NIMS
for the fruitful discussions, and  
to Seiko Matsumoto at NIMS
for the sample syntheses and X-ray diffraction measurements. 
We would like to thank Editage (www.editage.com) for English language editing.

\end{acknowledgments}

\newpage 

\begin{references}

\bibitem{Dender97}
D. C. Dender, P. R. Hammar, D. H. Reich, C. Broholm, and G. Aeppli, 
Direct Observation of Field-Induced Incommensurate Fluctuations
in a One-Dimensional $S = \frac{1}{2}$ Antiferromagnet, 
Phys. Rev. Lett. {\bf 79}, 1750 (1997).

\bibitem{Nagata76}
K. Nagata, 
Short range order effects on EPR frequencies in 
antiferromagnets with inequivalent $g$-tensors, 
J. Phys. Soc. Jpn. {\bf 40}, 1209 (1976).

\bibitem{Asano00}
T. Asano, H. Nojiri, Y. Inagaki, J. P. Boucher, T. Sakon, Y. Ajiro, and M. Motokawa, 
ESR Investigation on the Breather Mode and the 
Spinon-Breather Dynamical Crossover in Cu Benzoate,
Phys. Rev. Lett. {\bf 84}, 5880 (2000).

\bibitem{Nojiri06}
H. Nojiri, Y. Ajiro, T. Asano, and J. P. Boucher, 
Magnetic excitation of $S = \frac{1}{2}$ antiferromagnetic
spin chain Cu benzoate in high magnetic fields, 
New J. Phys. {\bf 8}, 218 (2006).

\bibitem{Kenzelmann04}
M. Kenzelmann, Y. Chen, C. Broholm, D. H. Reich, and Y. Qiu, 
Bound Spinons in an Antiferromagnetic $S = \frac{1}{2}$ Chain 
with a Staggered Field, 
Phys. Rev. Lett. {\bf 93}, 017204 (2004).

\bibitem{Kenzelmann05}
M. Kenzelmann, C. D. Batista, Y. Chen, C. Broholm, D. H. Reich, S. Park, and Y. Qiu, 
$S = \frac{1}{2}$ chain in a staggered field: High-energy bound-spinon state and 
the effects of a discrete lattice, 
Phys. Rev. B {\bf 71}, 094411 (2005).

\bibitem{Oshikawa99}
M. Oshikawa, K. Ueda, H. Aoki, A. Ochiai, and M. Kohgi, 
Field-Induced Gap Formation in Yb$_4$As$_3$, 
J. Phys. Soc. Jpn. {\bf 68}, 3181 (1999).

\bibitem{Kohgi01}
M. Kohgi, K. Iwasa, J. M. Mignot, B. F\aa k, P. Gegenwart, 
M. Lang, A. Ochiai, H. Aoki, and T. Suzuki, 
Staggered Field Effect on the One-Dimensional $S = \frac{1}{2}$ 
Antiferromagnet Yb$_4$As$_3$, 
Phys. Rev. Lett. {\bf 86}, 2439 (2001).

\bibitem{Feyerherm00}
R. Feyerherm, S. Abens, D. G\"unther, T. Ishida, M. Mei$\beta$ner, M. Meschke, 
T. Nogami, and M. Steiner, 
Magnetic-field induced gap and staggered susceptibility 
in the $S = \frac{1}{2}$ chain 
[PM $\cdot$ Cu(NO$_3$)$_2$ $\cdot$ (H$_2$O)$_2$]$_n$ (PM = pyrimidine), 
J. Phys.: Condens.Matter {\bf 12}, 8495 (2000).

\bibitem{Zvyagin04}
S. A. Zvyagin, A. K. Kolezhuk, J. Krzystek, and R. Feyerherm, 
Excitation Hierarchy of the Quantum Sine-Gordon Spin Chain in a Strong Magnetic Field, 
Phys. Rev. Lett {\bf 93}, 027201 (2004).

\bibitem{Wolter03}
A. U. B. Wolter, H. Rakoto, M. Costes, A. Honecker, W. Brenig, A. Kl\"umper, 
H.-H. Klauss, F. J. Litterst, R. Feyerherm, D. J\'erome, and S. S\"ullow, 
High-field magnetization study of the $S = \frac{1}{2}$ 
antiferromagnetic Heisenberg chain 
[PM Cu(NO$_3$)$_2$(H$_2$O)$_2$]$_n$ with a field-induced gap, 
Phys. Rev. B {\bf 68}, 220406(R) (2003).

\bibitem{Wolter05}
A. U. B. Wolter, P. Wzietek, S. S\"ullow, F. J. Litterst,
A. Honecker, W. Brenig, R. Feyerherm, and H.-H. Klauss,
Giant Spin Canting in the $S = \frac{1}{2}$ Antiferromagnetic Chain 
[CuPM(NO$_3$)$_2$(H$_2$O)$_2$]$_n$
Observed by $^{13}$C-NMR, 
Phys. Rev. Lett. {\bf 94}, 057204 (2005).

\bibitem{Morisaki07}
R. Morisaki, T. Ono, H. Tanaka, and H. Nojiri, 
Thermodynamic properties and elementary excitations 
in quantum sine-Gordon spin system KCuGaF$_6$, 
J. Phys. Soc. Jpn. {\bf 76}, 063706 (2007).

\bibitem{Umegaki09}
I. Umegaki, H. Tanaka, T. Ono, H. Uekusa, and H. Nojiri,
Elementary excitations of the $S = \frac{1}{2}$
one-dimensional antiferromagnet KCuGaF$_6$ 
in a magnetic field and quantum sine-Gordon model,
Phys. Rev. B  {\bf 79}, 184401 (2009).

\bibitem{Umegaki12}
I. Umegaki, H. Tanaka, T. Ono, M. Oshikawa, and K. Sakai,
Thermodynamic properties of quantum sine-Gordon spin chain system KCuGaF$_6$, 
Phys. Rev. B {\bf 85}, 144423 (2012).

\bibitem{Umegaki15}
I. Umegaki, H. Tanaka, N. Kurita, T. Ono, M. Laver, C. Niedermayer, C. R\"uegg,
S. Ohira-Kawamura, K. Nakajima, and K. Kakurai, 
Spinon, soliton, and breather in the spin-$\frac{1}{2}$ 
antiferromagnetic chain compound KCuGaF$_6$, 
Phys. Rev. B {\bf 92}, 174412 (2015).

\bibitem{Hinatsu14}
Y. Hinatsu and Y. Doi, 
Structural phase transition and antiferromagnetic transition of Tb$_3$RuO$_7$, 
J. Solid State Chem. {\bf 220}, 22 (2014).

\bibitem{Bontchev00}
R. P. Bontchev, A. J. Jacobson, M. M. Gospodinov, V. Skumryev, V. N. Popov, B. Lorenz, 
R. L. Meng, A. P. Litvinchuk, and M. N. Iliev, 
Crystal structure, electric and magnetic properties, and Raman spectroscopy of 
Gd$_3$RuO$_7$,
Phys. Rev. B  {\bf 62}, 12235 (2000).

\bibitem{Ishizawa08}
N. Ishizawa, K. Tateishi, S. Kondo, and T. Suwa, 
Structural phase transition of Gd$_3$RuO$_7$, 
Inorg. Chem. {\bf 47}, 558 (2008).

\bibitem{Groen87}
W. A. Groen, F. P. F. van Berkel, and D. J. W. IJdo, 
Trineodymium ruthenate(V). A Rietveld refinement of neutron powder diffraction data, 
Acta Crystallogr. C {\bf 43}, 2262 (1987).

\bibitem{Harada01}
D. Harada, Y. Hinatsu, and Y. Ishii, 
Studies on the magnetic and structural phase transitions of Nd$_3$RuO$_7$,
J. Phys.: Condens. Matter {\bf 13}, 10825(2001). 
 
\bibitem{Wakeshima10}
M. Wakeshima and Y. Hinatsu, 
Magnetic properties and structural transitions of orthorhombic fluorite-related 
compounds {\it Ln}$_3${\it M}O$_7$ 
({\it Ln} = rare earths, {\it M} = transition metals),
J. Solid State Chem. {\bf 183}, 2681 (2010).

\bibitem{Ida06}
T. Ida, K. Hiraga, H. Hibino, S. Oishi, D. du Boulay, and N. Ishizawa, 
A non-centrosymmetric polymorph of Gd$_3$RuO$_7$,
Acta Crystallogr. E {\bf 62}, i13 (2006).

\bibitem{Momma11}
K. Momma and F. Izumi, 
VESTA for three-dimensional visualization of 
crystal, volumetric and morphology data, 
J. Appl. Cryst. {\bf 44}, 1272 (2011).

\bibitem{Harada02}
D. Harada and Y. Hinatsu, 
Magnetic and calorimetric studies on one-dimensional 
{\it Ln}$_3$RuO$_7$ ({\it Ln} = Pr, Gd), 
J Solid State Chem. {\bf 164}, 163 (2002).

\bibitem{hrpt}
P. Fischer, G. Frey, M. Koch, M. Koennecke, V. Pomjakushin, J. Schefer, R. Thut,
N. Schlumpf, R. Buerge, U. Greuter, S. Bondt, and E. Berruyer, 
High-resolution powder diffractometer HRPT for thermal neutrons at SINQ, 
Physica B {\bf 276-278}, 146 (2000); 
[http://sinq.web.psi.ch/hrpt].

\bibitem{Campbell06}
B. J. Campbell, H. T. Stokes, D. E. Tanner, and D. M. Hatch, 
ISODISPLACE: An internet tool for exploring structural distortions, 
J. Appl. Cryst. {\bf 39}, 607 (2006);
[https://stokes.byu.edu/iso/isodistort.php].

\bibitem{Rodriguez93}
J. Rodriguez-Carvajal, 
Recent advances in magnetic structure determination by neutron powder diffraction, 
Physica B  {\bf 192}, 55 (1993); 
[http://www.ill.eu/sites/fullprof/]. 


\bibitem{Das17}
M. Das, S. Roy, and P. Mandal, 
Giant reversible magnetocaloric effect in a multiferroic 
GdFeO$_3$ single crystal,
Phys. Rev. B {\bf  96}, 174405 (2017). 

\bibitem{Boutahar17}
A. Boutahar, R. Mouba, E. K. Hlil, H. Lassri, E. Lorenzo, 
Large reversible magnetocaloric effect in antiferromagnetic Ho$_2$O$_3$ powders,
Sci. Rep. {\bf  7}, 13904 (2017).

\bibitem{Hase21}
M. Hase, N. Tsujii, H. S. Suzuki, O. Sakai, and H. Mamiya, 
Magnetic properties of oxides with high concentrations of 
rare-earth elements R$_6$AO$_{12}$ 
(R = rare-earth element, A = Mo or W), 
J. Magn. Magn. Mater. {\bf 523}, 167539 (2021).

\bibitem{Paul16}
R. Paul, T. Paramanik, K. Das, P. Sen, B. Satpati, I. Das, 
Magnetocaloric effect  at cryogenic temperature in gadolinium oxide nanotubes, 
J. Magn. Magn. Mater. {\bf  417}, 182 (2016).

\bibitem{Hase20}
M. Hase, N. Tsujii, and H. Mamiya, 
Magnetocaloric effect in the double perovskites 
Sr$_2${\it R}RuO$_6$ ({\it R} = Dy and Tb), 
J. Jpn. Soc. Powder powder metallurgy {\bf 67}, 182 (2020) 

\bibitem{Cai17}
Y. Q. Cai, Y. Y. Jiao, Q. Cui, J. W. Cai, Y. Li, B. S. Wang, M. T. Fern\'andez-D\'iaz, 
M. A. McCuire, J.-Q. Yan, J. A. Alonso, J.-G. Cheng, 
Giant reversible magnetocaloric effect in the pyrochlore Er$_2$Mn$_2$O$_7$ 
due to a cooperative two-sublattice ferromagnetic order, 
Phys. Rev. Mater. {\bf  1}, 064408 (2017).

\bibitem{Jin11}
J.-L. Jin, X.-Q. Zhang, G.-K. Li, Z.-H. Cheng, L. Zheng, and Y. Lu, 
Giant anisotropy of magnetocaloric effect in TbMnO$_3$ single crystals,
Phys. Rev. B  {\bf 83}, 184431 (2011). 

\bibitem{Omote19}
H. Omote, S. Watanabe, K. Matsumoto, I. Gilmutdinov, A. Kiiamov, D. Tayurskii, 
Magnetocaloric effect in single crystal GdTiO$_3$, 
Cryogenics {\bf 101}, 58 (2019).

\bibitem{Su13}
Y. Su, Y. Sui, J.-G. Cheng, J.-S. Zhou, X. Wang, Y. Wang, and J. B. Goodenough, 
Critical behavior of ferromagnetic perovskites RTiO$_3$ (R = Dy, Ho, Er, Tm, Yb) 
by magnetocaloric measurements, 
Phys. Rev. B {\bf 87}, 195102 (2013).

\bibitem{Su11}
Y. Su, Y. Sui, J. Cheng, X. Wang, Y. Wang, W. Liu, and X. Liu, 
Large reversible magnetocaloric effect in HoTiO$_3$ single crystal, 
J. Appl. Phys. {\bf 110}, 083912 (2011).

\bibitem{Su12a}
Y. Su, Y. Sui, X. Wang, J. Cheng, Y. Wang, W. Liu, and X. Liu, 
Large magnetocaloric properties in single-crystal dysprosium titanate, 
Mater. Lett. {\bf 72}, 15 (2012).

\bibitem{Su12b}
Y. Su, Y. Sui, J. Cheng, X. Wang, Y. Wang, P. Liu, and J. Tang, 
Large reversible magnetocaloric effect in TmTiO$_3$ single crystal, 
J. Appl. Phys. {\bf 111}, 07A925 (2012).

\bibitem{Franco18}
As a review article, 
V. Franco, J. S. Bl\'azquez, J. J. Ipus, J. Y. Law, 
L. M. Moreno-Ram\'irez, and A. Conde, 
Magnetocaloric effect: 
From materials research to refrigeration devices, 
Progr. Mater. Sci. {\bf  93}, 112 (2018).


\bibitem{Vilminot03}
S. Vilminot, M. Richard-Plouet, G. Andr\'e, D. Swierczynski, M. Guillot, 
F. Bour\'ee-Vigneron, and M. Drillon, 
Magnetic structure and properties of Cu$_3$(OH)$_4$SO$_4$
made of triple chains of spins $s = \frac{1}{2}$, 
J. Solid State Chem. {\bf 170}, 255 (2003).

\bibitem{Vilminot07}
S. Vilminot, G. Andr\'e, F. Bour\'ee-Vigneron, M. Richard-Plouet, and M. Kurmoo,
Magnetic properties and magnetic structures of 
Cu$_3$(OH)$_4$XO$_4$, X = Se or S: 
Cycloidal versus collinear antiferromagnetic structure,
Inorg. Chem. {\bf 46}, 10079 (2007).

\bibitem{Hase15}
M. Hase, H. Kuroe, V. Yu. Pomjakushin, L. Keller, R. Tamura, and N. Terada 
Y. Matsushita, A. D\"onni, and T. Sekine, 
Magnetic structure of the spin-$\frac{1}{2}$ frustrated quasi-one-dimensional antiferromagnet 
Cu$_3$Mo$_2$O$_9$: Appearance of a partially disordered state, 
Phys. Rev B {\bf  92}, 054425 (2015).

\bibitem{Mekata76}
M. Mekata, 
Antiferro-ferrimagnetic transition in triangular Ising lattice, 
J. Phys. Soc. Jpn. {\bf 42}, 76 (1977).

\bibitem{Mekata93}
M. Mekata, N. Yaguchi, T. Takagi, T. Sugino, S. Mitsuda, H. Yoshizawa, 
N. Hosoito, and T. Shinjo, 
Successive magnetic ordering in CuFeO$_2$ 
-A new type of partially disordered phase in a triangular lattice antiferromagnet, 
J. Phys. Soc. Jpn. {\bf 62}, 4474 (1993).

\bibitem{Matsuda12}
M. Matsuda, C. de la Cruz, H. Yoshida, M. Isobe, R. S. Fishman, 
Partially disordered state and spin-lattice coupling in an $S = \frac{3}{2}$
 triangular lattice antiferromagnet Ag$_2$CrO$_2$, 
Phys. Rev B {\bf 85}, 144407 (2012).

\bibitem{Stewart04}
J. R. Stewart, G. Ehlers, A. S. Wills, S. T. Bramwell, and J. S. Gardner,
Phase transitions, partial disorder and multi-$k$ structures in 
Gd$_2$Ti$_2$O$_7$, 
J. Phys.: Condens. Matter {\bf 16}, L321 (2004).

\bibitem{Matsuda08}
M. Matsuda, J.-H. Chung, S. Park, T. J. Sato, K. Matsuno, H. Aruga Katori, 
H. Takagi, K. Kakurai, K. Kamazawa, Y. Tsunoda, I. Kagomiya, 
C. L. Henley, and S.-H. Lee, 
Frustrated minority spins in GeNi$_2$O$_4$, 
Europhys. Lett. {\bf 82}, 37006 (2008).

\end{references}

\end{document}